\documentclass{PoS}

\def\be{\begin{eqnarray} &&} 
\def\nonu{\nonumber \\ &&} 
\def\ee{\end{eqnarray}} 
\newcommand{\bm}[1] {\mbox{\boldmath{$#1$}}}
\newcommand{\blf}[1]{\bf  {\tilde #1}}
\def\sumint{\int \! \!\ \! \! \! \! \!\ \! \! \!\! \!\sum}

\def\bq{{\bf q}}

\newcommand{\la}{\langle\,}
\newcommand{\ra}{\,\rangle}

\def \bkappa {{\mbox{\boldmath$\kappa$}}}

\title{Neutron Parton Structure And The Light-Front Spectral Function Of 3He}

\ShortTitle{Neutron Parton Structure And The Light-front Spectral Function Of 3He}

\author{\speaker{Emanuele Pace}%
        \\
       Universita' degli Studi di Roma Tor Vergata, Dipartimento di Fisica, and INFN, Sezione di Roma Tor Vergata, Italy,
       Via della Ricerca Scientifica 1, 00133 Rome, Italy,\\
       E-mail: \email{pace@roma2.infn.it}}
\author{Alessio Del Dotto\\
       Istituto  Nazionale di Fisica Nucleare, Sezione di Roma,  P.le A. Moro 2, 00185 Rome, Italy and
       University of South Carolina, Columbia, SC 29208, USA}
\author{Leonid Kaptari\\
      Bogoliubov Lab. Theor. Phys., 141980, JINR, Dubna, Russia}
\author{Matteo Rinaldi\\
Instituto de Fisica Corpuscular (CSIC-Universitat de Valencia), Parc Cientific UV,
C/ Catedratico Jose Beltran 2, E-46980 Paterna (Valencia), Spain}
\author{Giovanni Salm\`e\\
Istituto  Nazionale di Fisica Nucleare, Sezione di Roma,  P.le A. Moro 2, 00185 Rome, Italy}
\author{Sergio Scopetta\\
       Dipartimento di Fisica e Geologia, Universit\`a degli Studi di Perugia and INFN, Sezione di Perugia, Via Alessandro Pascoli,
06123 Perugia, Italy
}

\abstract{Semi-inclusive deep inelastic electron scattering off polarized $^3He$ is studied in a non-relativistic framework, using a distorted spin-dependent spectral function for $^3He$  to take care of the final state interaction between the observed pion and the remnant. A simple procedure is shown to be valuable for the extraction of Sivers and Collins asymmetries for the neutron from the corresponding asymmetries for $^3He$ .To extend this study in a relativistic framework, a novel approach for a Poincar\'e covariant description of nuclear dynamics  is presented, based on the
light-front Hamiltonian dynamics. 
The key quantity is the light-front spectral function, where both 
normalization and momentum sum rule can be satisfied at the same time. 
Preliminary results are discussed for an initial analysis of the role of relativity in the EMC effect in $^3He$, and
 the generalization of the  procedure for the extraction of neutron asymmetries  within the light-front dynamics is outlined.}

\FullConference{The 26th International Nuclear Physics Conference\\
	 11-16 September, 2016\\
	 Adelaide, Australia}

\begin{document}

\section{Introduction}
\label{intro}

Semi-inclusive deep inelastic (SIDIS) processes  where a fast meson is detected are an important tool for the knowledge of the internal hadron dynamics. Indeed, the detected
meson likely originates from the fragmentation of the quark which absorbed the virtual photon and  opens a valuable window on the motion of quarks inside the parent nucleon, before the
interaction with the photon. Hence,  through  SIDIS reactions (see, e.g., \cite{Qian,SIDIS,06010}) one can access the
transverse-momentum-dependent parton distributions (TMDs)  of nucleons (see, e.g., Ref. \cite{BARONE}). Neutron targets are not available, but,
   within a non-relativistic approach  which include the final state interaction (FSI) through a distorted spin-dependent spectral function (SF), it has been shown the actual possibility to get information on the neutron structure from SIDIS experiments on 
$^3He$ \cite{Kaptari,DelDotto,Kaptari1}. 

 For a relativistic description of few-body nuclei, we adopt a Poincar\'e covariant spin-dependent SF \cite{DPSS}, built up within the light-front Hamiltonian dynamics  (LFHD) for an interacting system with a fixed number of on-mass-shell constituents  (see, e.g., \cite{KP}). The LFHD has
a subgroup structure of the light-front (LF) boosts (with a separation of the intrinsic motion from the global one) and allows one to give a  fully Poincar\'e covariant description of deep inelastic scattering (DIS), SIDIS and deeply virtual Compton scattering.
Furthermore, within the LFHD and
using the Bakamijan-Thomas (BT) construction of the Poincar\'e generators \cite{Baka}  one can  take advantage of the whole successful non-relativistic (NR) phenomenology  that has been 
developed for the nuclear interaction. 
A distinct feature of our approach is the ability to
implement macrocausality or cluster separability, namely the expected property that if a system is separated into disjoint subsystems by a sufficiently large spacelike separation, then the subsystems behave as independent systems.

 In Section 2  the procedure to obtain information on the neutron Collins and Sivers asymmetries from SIDIS experiments on 
$^3He$  is discussed.
In Section 3 the LF spin-dependent (SD) SF obtained from the LF wave functions for  two- and  three-nucleon systems is described, and the generalization to the LF dynamics of our procedure for the extraction of neutron asymmetries  is outlined. In Section 4 the LF SF is applied to study the role of relativity for the EMC effect in $^3He$ and preliminary results are presented.
In Section 5 conclusions and perspectives  are drawn.

\section{Extraction of neutron asymmetries from SIDIS experiments off $^3He$}

\vspace{-1mm}
The Collins and Sivers asymmetries, $A_{3}^{C(S)}$, can be expressed as follows 
 \vspace{-1mm}
 \begin{eqnarray}
A_{3}^{C(S)} =
\frac
{
\int_{x}^A 
d\alpha
\left[
\Delta \sigma_{C(S)}^n\left (x/\alpha ,Q^2
\right  ) 
{{f^{\perp}_n(\alpha ,Q^2)}}+
2\Delta 
\sigma_{C(S)}^p\left (x / \alpha ,Q^2 \right ) 
{{f^{\perp}_p(\alpha ,Q^2)}} \right]
}
{\int d\alpha\left[
 \sigma^n\left (x/\alpha ,Q^2 \right ) {{f^{~}_n(\alpha ,Q^2)}}+
2\sigma^p\left (x/\alpha ,Q^2 \right ) {{f^{~}_p(\alpha ,Q^2)}} \right]}
\label{asi}
\end{eqnarray}
\vspace{-1mm}
in terms of the light-cone unpolarized,  {{$f^{~}_N$}} , and transverse,  {{$f_N^\perp$}} , momentum distributions (md)
\begin{eqnarray}
\hspace{-2mm}{{f_N^{~(\perp)}(\alpha,Q^2)}} =
\int dE \int_{p_{m}(\alpha,Q^2)}^{p_{M}(\alpha,Q^2)}
 {m_N \over E_N}
{{{P}_N^{~(\perp)}(E, {\bf p})}} \, \delta \left ( \alpha - {p\cdot q \over m_N \nu} \right )
\, \theta \left(W_Y^2- \left(m_N+m_\pi \right )^2 \right ) d^3 {\bf{p}}
\label{md}
\end{eqnarray}
with $W_Y$ the invariant mass of the debris Y, which hadronizes in a nucleon and, at least, one pseudoscalar meson. The quantities $\Delta \sigma_{C(S)}^{N}$
and $\sigma^{N}$ in Eq. (\ref{asi}) are
related to the structure of the bound nucleon
\begin{eqnarray}
\hspace{-5mm}\Delta \sigma_{C}^N\left(x,Q^2 \right  )
& = &
{ {1 -y \over 1-y-y^2/2}}
\nonumber \\
& \times &
\sum_q e_q^2
\int  d^2 {\bkappa_T}
d^2 {\bf k}_T \delta^2 ( {\bf k}_T + {\bf q}_T  - \bkappa_T )
 {{\bf \hat{P}}_{h\,\perp} \cdot {\bkappa_T} \over m_h}
 h_1^{q,N} (x, {\bf k}_T^2 )
 H_1^{\perp q,h} (z, (z {\bkappa_T})^2 )~,
\label{dcoll}
\end{eqnarray}
\be
\hspace{-5mm}\Delta \sigma_{S}^N\left (x,Q^2 \right  )
=
 \sum_q e_q^2
\int d^2 {\bkappa_T}
d^2 {\bf k}_T
\delta^2 ( {\bf k}_T + {\bf q}_T  - {\bkappa_T} )
{ {\bf \hat{P}}_{h\,\perp} \cdot {\bf{k}_T} \over m_N}
f_{1T}^{\perp q,N} (x, {\bf{k}}_T^2 )
  D_1^{q,h} (z, (z \bkappa_T)^2 )~,
\label{dsiv}
\ee
\be
\sigma^N\left (x,Q^2 , z\right  )
=
\sum_q e_q^2
\int d^2 {\bkappa_T} d^2 {\bf k}_T
\delta^2 ( {\bf k}_T + {\bf q}_T  - {\bkappa_T} )
f_1^{q,N} (x,{\bf k}_T^2 )
D_1^{q,h}  (z, (z {\bkappa_T})^2 )~,
\label{unpol}
\ee
where $z = E_h/ \nu$ and
models for the parton distributions $ h_1^{q,N}$, $f_{1T}^{\perp q,N}$,
${{f_1^{q,N}}}$, and for the fragmentation functions $H_1^{\perp q,h}$,
${{D_1^{q,h}}}$ were used (see Ref. \cite{mio}).
In Eq. (\ref{md}) ${P}_N^{~}(E, {\bf p})$ is the unpolarized SF (see \cite{cps}), while  ${P}^{\, \perp}_N(E,{\bf p})$  is the transverse SF 
\be
{P}^{\, \perp}_N(E,{\bf p})= \Re e \left \{ {P}^{N \, \frac12 -\frac12}_{\frac12 -\frac12}(E,{\bf p}) 
+ {P}^{N \, -\frac12 \frac12}_{\frac12 -\frac12}(E,{\bf p}) \right \} \quad\quad.
\label{trspectr}\ee
In Ref. \cite{Kaptari1} the matrix elements of a distorted SD SF
which includes a generalized eikonal approximation (GEA)  to take care of the FSI  in a NR approach were introduced
\be
{P}^{N \,MM'}_{\lambda\lambda'}(E,{\bf p})=
\sum_{f_{23}} 
\sum \! \!\! \!\! \!\! \!\int_{~\epsilon^*_{23}}\rho\left(
\epsilon^*_{23}\right)\,
{ \tilde {\cal O}}_{\lambda\lambda'}^{N \, MM' \, f_{23}}
(~\epsilon^*_{23},{\bf p})
\,
{ \delta\left(  E+ M_3-m_N-M^*_{23}\right)}~
~,
\label{spectrg}
\ee
with $\hspace{7mm}$
${ \tilde {\cal O}}_{\lambda\lambda'}^{N \,M \, M' \, f_{23}}
(\epsilon^*_{23},{\bf p})=
\la \lambda, {\bf p};
{{\hat S_{Gl}}} \phi_{\epsilon_{23}^*}^{f_{23}} 
| 
\Psi_3^{M}\ra
\la \Psi_3^{M'}| \lambda', {\bf p}; {{\hat S_{Gl}}}
\phi_{\epsilon_{23}^*}^{f_{23}}
\ra.$

The spin components $M$, $M'$ and $\lambda$, $\lambda'$ are defined with
respect to the direction of $\hat\bq$. 
The operator
$\, {{\hat S_{Gl}}} 
({\bf r}_1,{\bf r}_2,{\bf r}_3)=
\prod_{i=2,3}\bigl[1-\theta(z_i-z_1)
{{\Gamma}}({\bf b}_1-{\bf b}_i,{ z}_1-{z}_i)
\bigr]
$ 
is a {Glauber} operator  
which takes care of  hadronization and FSI. The model of Ref. \cite{Kope1} for the (generalized) {{profile function} ${{\Gamma({{\bf b}},z)}}$,
already successfully applied to $^2H(e,e'p)X$ \cite{Ciofi}, is adopted.

In Ref. \cite{mio}, using the NR SF of 
Ref. \cite{cps} and within the plane wave impulse approximation (IA), i.e. no interaction between the measured fast $\pi$, the remnant debris and the interacting two-nucleon recoiling system,
it was shown that the formula \cite{neutr}
\be
  A_n \simeq {1 \over 
{{p_n}} d_n} \left 
( {{A^{exp}_3}} - 2 
{{p_p}} d_p
{{A^{exp}_p}} \right )~, \quad 
\label{extrac}
\ee
already widely used to extract neutron asymmetries in DIS from experiments on $^3He$, works also in SIDIS, both for the Collins and Sivers single spin asymmetries. Nuclear effects are hidden in the effective polarizations (EP) $p_p=-0.024$ and $p_n=0.878$ and in the dilution factors, $d_{p(n)}$.

To investigate whether the formula (\ref{extrac}) can be safely applied even in presence of the FSI, the GEA distorted spin-dependent SF was adopted in Ref. \cite{Kaptari1}.
While {{$P^{IA}$}} 
 depends on ground state properties, 
{{$P^{FSI}$}} 
is  process dependent, since the Glauber operator depend on the kinematics of the process. Then for each  experimental point ($x, Q^2...$) a different 
$P^{FSI}$ has to be evaluated !

The SFs ${P}^{IA}$
and 
${P}^{FSI}$, as well as the 
light-cone md  $f_N^{IA}$ and  $f_N^{FSI}$ 
can be very different  (see Fig. 1)
  and therefore
FSI's have  a strong effect on the 
SIDIS cross sections.
\begin{figure}[h]
\vspace{-0.8cm}
\includegraphics[width=0.45\textwidth]{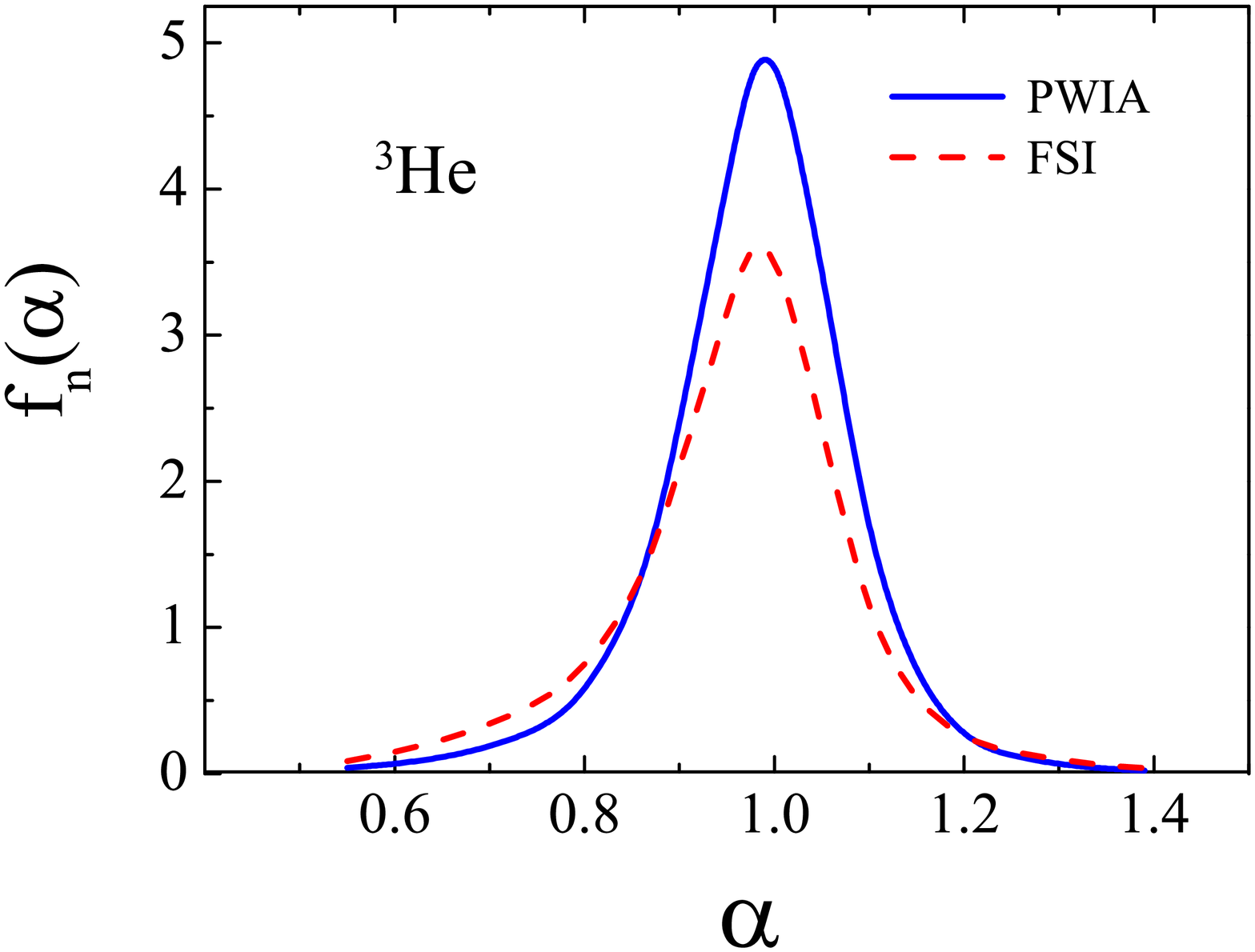}
\includegraphics[width=0.45\textwidth]{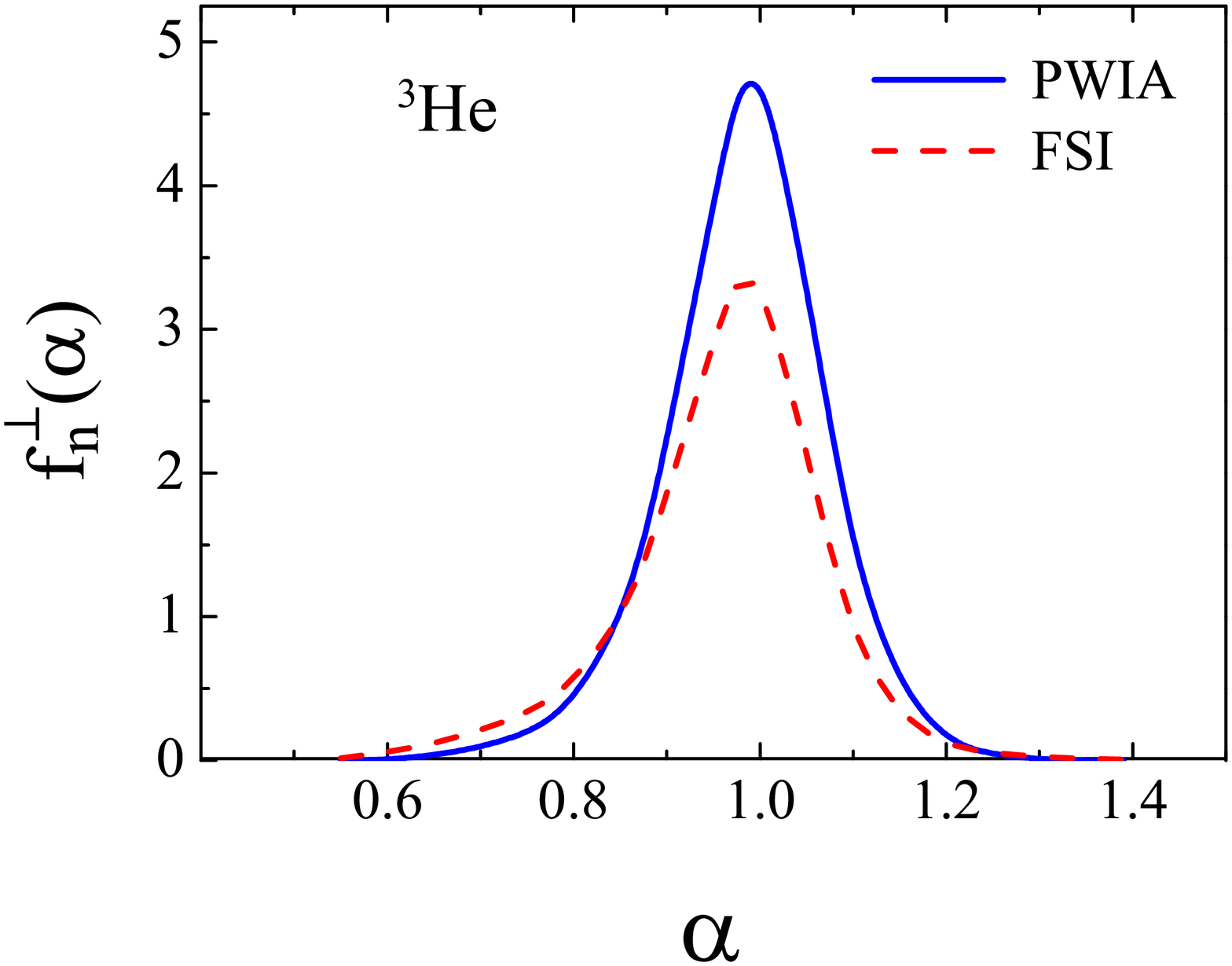}
\vspace{-0.3cm}
\caption{Neutron unpolarized and transversely polarized distributions in $^3He$ in IA (full lines) and with FSI (dashed lines) for the initial electron energy $\cal{E}$= 8.8 GeV 
and $Q^2 = 5.73 ~(GeV/c)^2$ (preliminary results).}
\label{distr}
\end{figure}
However,  including the FSI the  md  {{$f_N$}} and  {{$f_N^\perp$}} change in the same way and in asymmetries the md appear  both in the numerator and in the denominator. Furthermore, while FSI's change effective polarizations $p_{p(n)}$ 
by 10-15 \%,
{{it occurs that effects of GEA-FSI in the dilution factors and in the
EP compensate each other
to a large extent: i.e., the products ~
{{$p^{FSI}_{p(n)}~d^{FSI}_{p(n)}$}}  and ~
{{$p^{IA}_{p(n)}~d^{IA}_{p(n)}$}}
are essentially the same
 \cite{DelDotto}. Then the {{usual extraction}} of Eq. (\ref{extrac}) is safe, as shown at ${\cal E}=$ 8.8 GeV in Fig. \ref{asymm}.
}}
\begin{figure}[h]
\includegraphics[width=14.cm]{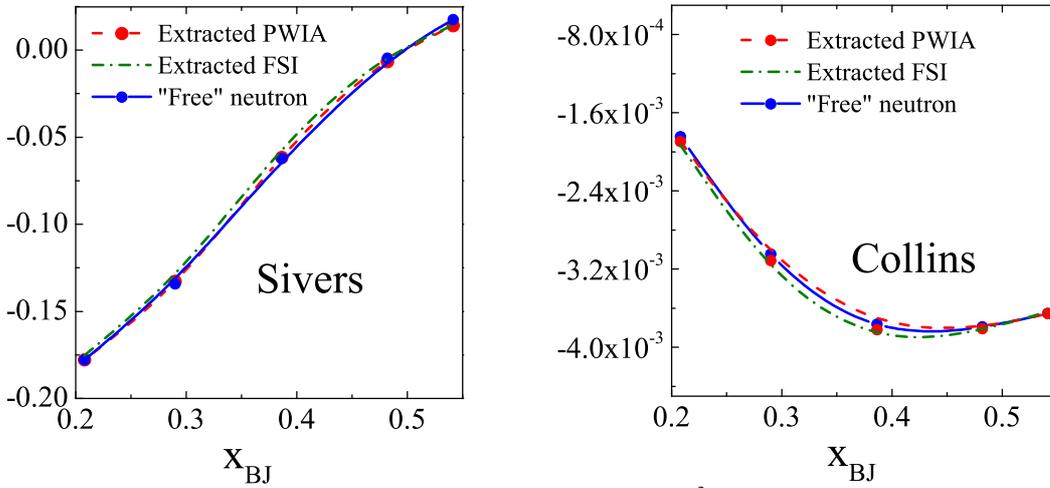}
\vskip -5.mm
\caption{Neutron asymmetries extracted 
(Eq. (2.7))
from the $^3He$ Sivers (left panel) and Collins (right panel) 
asymmetries, with and without FSI, in the actual kinematics of JLab
\cite{SIDIS} (preliminary results, see \cite{Kaptari1}).
}
\label{asymm}
\end{figure}

\section{Light-Front Dynamics and the Light-Front Spectral Function}
An explicit construction of the 10 Poincar\'e generators that fulfills the proper commutation rules in presence of interactions was given by Bakamjian and Thomas \cite{Baka} : i) only   the mass operator, M, contains the interaction, and ii) it generates the dependence upon the interaction of the three dynamical generators in LFHD, namely $P^-$ and the  transverse rotations $\vec F_\perp$.
M
is obtained adding to the free mass $M_0$ of the system an interaction $V$. There are two possibilities: $M^2 = M_0^2 + U$ (then for  two particles one can easily embed the NR phenomenology) or $M = M_0 + V$.
The interaction, $U$ or $V$, must commute with all the kinematical generators, and with the non interacting spin. Then it has to be invariant for translations and rotations, as in the NR case.

For the three-body case the mass operator is
$M_{BT}(123)= {{M_0(123)}}+ V^{BT}$,
where
$M_0(123)= \sqrt{m^2 +k^2_1}+\sqrt{m^2 +k^2_2}+\sqrt{m^2 +k^2_3}$~
is the free mass operator,	$V^{BT}$	a BT two-body and  three-body force, and $k_i~(i=1,2,3)$ are intrinsic momenta with	
${\bf k}_1 +{\bf  k}_2 +{\bf k}_3=0$
 \cite{KP}.

The NR mass operator is written as
$M^{NR}=3m + \sum_{i=1,3} {k^2_i / 2m}
+V^{NR}_{12}+V^{NR}_{23}+V^{NR}_{31}+V^{NR}_{123}$
and must obey  the commutation rules proper of the Galilean group, leading to translational and rotational invariance.
 Those properties are analogous to the ones in the BT construction. This allows us to consider the standard NR mass operator as a sensible BT mass operator, and embed it in a Poincar\'e covariant approach:
$M_{BT}(123) \sim M^{NR}~$.

 To obtain  within the LFHD a Poincar\'e-covariant 
 spin-dependent 
 SF for a three-particle system in the bound state $|\Psi_{0};S, T_z \rangle$, eigenstate of the mass operator $M_{BT}(123)$  and polarized along
$\vec{S}$,
 let us use the LF overlaps
  $_{LF}\langle \tau_{S},T_{S};\alpha,\epsilon ;J_{z}J;\tau\sigma,\tilde{\bm \kappa}|\Psi_{0}; S, T_z\rangle$ in place of  their NR  counterparts in the definition of the SF. 
 The state $_{LF}\langle \tau_{S},T_{S};\alpha,\epsilon ;J_{z}J;\tau\sigma,\tilde{\bm \kappa}|$ is the tensor product of a plane wave for the knocked-out constituent (say particle 1) with intrinsic momentum
$\tilde{\bm \kappa}$, and a fully interacting intrinsic state for the spectator system (say particles 2 and 3), with energy $\epsilon$, \emph{all moving in the intrinsic reference frame 
 of the cluster (1,23)}.
When applications to DIS or SIDIS processes are concerned, the issue of macrocausality has to be considered, i.e., if
the subsystems which compose a system are brought far apart, 
the Poincar\'e generators of the system have to become the sum 
of the Poincar\'e generators corresponding to the subsystems in 
which the system is asymptotically separated. 
The packing operators \cite{KP}, which make it 
possible to include the macrocausality in the bound 
state, are not considered in 
the present approximation.
 However, we implement macrocausality in the tensor 
product of a plane wave for the knocked-out constituent times 
a fully interacting intrinsic state for the spectator pair. 
Then, the LF spin-dependent SF for 
 the three-nucleon system ($^3He$ or $^3H$) 
   is \cite{DPSS}
\be
 \hspace{-0.4cm}{{{\cal {P}}^{\tau}_{\sigma'\sigma}(\xi,{\bm \kappa}_\perp,\kappa^-,S)}
= 
\left|{\partial \kappa^+\over \partial \xi}\right|
~\sumint  d\epsilon~\rho(\epsilon) ~
\delta\left( \kappa^- -M_3+{M^2_S +|{\bm \kappa}_\perp|^2 \over (1-\xi)M_3}
\right)} 
~
~\times
\nonu
\hspace{-0.4cm} 
\sum_{J J_{z}\alpha}\sum_{T_{S}\tau_{S} } ~
_{LF}\langle  \tau_{S},T_{S} , 
\alpha,\epsilon; J J_{z}; \tau\sigma',\tilde{\bm \kappa}|\Psi_{0}; S,T_z
\rangle
  ~\langle S,T_z;
\Psi_0|\tilde{\bm \kappa},\sigma\tau; J J_{z}; 
\epsilon, \alpha, T_{S}, \tau_{S}\rangle_{LF} 
\label{LFspf}
\ee
where $\tau= \pm 1 /2$, $M_3$ is  the nucleus mass, $\rho(\epsilon)$  the  density of the
 two-nucleon eigenstates  
($\rho(\epsilon)  =  \sqrt{\epsilon ~ m} ~ m/2$ for the two-body continuum states and
$\rho(\epsilon)  = 1$ for the deuteron  bound state),
 $J$
  the spin and $T_{S}$
  the isospin  of the two-body state, 
 $\alpha$ the set of quantum numbers needed to completely specify this
  eigenstate,  and
  { $M_S=2\sqrt{m^2 +m\epsilon}$}  its mass. From ~$\xi, M_S, {\bm \kappa}_\perp$~ one can define
~{$\kappa^+=\xi{\cal M}_{0}(1,23)$,  where ${\cal M}_0(1,23)$ is the free mass of the cluster (1,23)
\be
 {\cal M}^2_{0}(1,23)={m^2 +|{\bm \kappa}_\perp|^2 \over \xi}+
{M^2_S +|{\bm \kappa}_\perp|^2 \over (1-\xi)} \quad .
\ee
  
The overlap 
$_{LF}\langle \tau_{S},T_{S};\alpha,\epsilon ;J_{z}J;\tau\sigma,\tilde{{\bm \kappa}}|\Psi_{0}; S, T_z\rangle$ is  defined as follows \cite{DPSS}
\be
\hspace{-6mm}_{{LF}}\langle  \tau_{S},T_{S} , 
\alpha,\epsilon; J_{z} J; \tau\sigma,{{\tilde{\bm \kappa}}}|\Psi_{0}; S, T_z
 \rangle = 
\sum_{\tau_2,\tau_3} \int d{\bf  k}_{23} 
\sum_{\sigma'_1}~ D^{{1 \over 2}} [{\cal R}_M 
(\tilde{\bm k} )]_{\sigma\sigma'_1}
~\sqrt{ 
  {\kappa^+ E_{23} \over k^+ E_S}}~\sqrt{(2 \pi)^3~2E({\bf k})}\times \nonu
\hspace{-6mm}
  ~\sum_{\sigma''_2,\sigma''_3}\sum_{\sigma'_2,\sigma'_3}
~\sum_{\sigma_2}~
 D^{{1 \over 2}} [{R}^\dagger_M{ ({\blf k}_{23} )}]_{\sigma''_2\sigma_2}~
 D^{{1 \over 2}} [{R}_M {({\blf k}_{2} )}]_{\sigma_2\sigma'_2}
 ~\sum_{\sigma_3}~ D^{{1 \over 2}} [{R}^\dagger_M{(-{\blf k}_{23} )}]_{\sigma''_3\sigma_3}~
 D^{{1 \over 2}} [{R}_M {({\blf k}_{3} )}]_{\sigma_3\sigma'_3}
\nonu \times ~ 
~~ _{{ {IF}}}\langle  \tau_S, T_{S},
\alpha,\epsilon; J_{z} J |{\bf k}_{23};\sigma"_2,\sigma"_3;\tau_2,\tau_3
\rangle 
\langle \tau_3,\tau_2,\tau; \sigma'_3, \sigma'_2, \sigma'_1;
{{\bf k}},{\bf k}_{23}| \Psi_{0}; S,T_z \rangle_{{{IF}}} \quad ,
\label{overl}
 \ee
where ${\bf k}_{23}$ is the intrinsic momentum of the (23) pair, ${\bf k}$ is the intrinsic nucleon momentum in the (123) system
 ({${\bf k}_\perp={\bm \kappa}_\perp$, since we choose the $^3He$
 transverse momentum ${\bf P}_\perp=0$}),
  {$k^+ = \xi~ M_0(123)= 
 \kappa^+ ~M_0(123)/ {\cal M}_0(1,23)$, with $M_0(123)$ the free mass of the three-particle system
  \be
  ~~~~~~~M^2_0(123)={m^2 +|{\bm k}_\perp|^2 \over \xi}+{M^2_{23} +
  |{\bm k}_\perp|^2 \over (1-\xi)}
  \ee
  and $~M^2_{23}= 4 (m^2 +|{\bf k}_{23}|^2)$
  the mass of the spectator pair without interaction ! 
In Eq. (\ref{overl}) one has  {$k_z= { 1\over 2} ~\left[k^+ -{(m^2+|{\bm \kappa}_\perp|^2 ) / k^+}
 \right]$, $E_{23}=\sqrt{M^2_{23}+|{\bf k}|^2}$ and $E_S=\sqrt{M^2_S+|{\bm
 \kappa}|^2}$}. 
 Furthermore 
{{ $D^{s}_{\sigma,\sigma'}(R^\dagger_{M}(\blf k))$}} 
is the Wigner function,  needed for  coupling  angular momenta in LFHD,
and the Melosh rotation {{ $R_{M}(\blf k)$}} is the rotation between the rest frames of the 
 particle reached through 
a LF boost or a canonical, rotationless boost \cite{KP}.
In our calculations, we identify the instant form (IF) overlaps of Eq. (\ref{overl}) with the NR wave functions for the two-nucleon and the three-nucleon \cite{pisa} systems,  corresponding to the NN interaction AV18 \cite{AV18}. 

We are presently planning to test our extraction procedure of neutron asymmetries from $^3He$ asymmetries using the LF SF and  including in our LF description  the FSI between the jet produced from the hadronizing quark and the two-nucleon spectator system through
an extension to the LF framework of the GEA of Refs. \cite{Kope1,Ciofi}, as we did in the NR case
\cite{Kaptari,Kaptari1}.

\section{Light-front momentum distribution and
preliminary results for the EMC effect}
  From the LF  SF one can obtain the  momentum distribution {{$f^A_{p(n)}(z)$}} 
 \be
 f^A_{\tau}(z)  = \int_0^1 d\xi \int d  {\bm \kappa}_\perp\int d\kappa^- ~{1 \over 2 (2 \pi)^3 \kappa^+} ~
 Tr \left[{{{\cal P}^{\tau}(\xi,{\bf k}_\perp, \kappa^-,S)}}\right]~ 
 \delta\left(z - {\xi M_A\over m} \right )
 \ee
that naturally fulfills 
 normalization and momentum sum rule
\vspace{-1mm}
\be
{{\int_0^{M_A/m} dz~f^A_{\tau}(z)=1 }} \quad \quad \quad
{{MSR}={1 \over A}\int_0^{M_A/m} dz~z~\left [Z f^A_{p}(z) + (A-Z) f^A_{n}(z)\right]
 ={ M_A\over A~m} }
\ee
because of the symmetry of the three-body bound state  
(see \cite{DPSS}).
 To investigate whether the LF SF can affect the EMC effect, we first evaluated the nuclear structure function { {$ F^A_2(x)$}} ($x=Q^2/2m\nu$)  as a convolution of the nuclear SF and of the nucleon structure function for the proton and the neutron.   
Then we obtained the ratios
\be
{{
R^A_2(x)={~ F^A_2(x)\over Z~F^p_2(x)+(A-Z)~F^n_2(x)}}}
\ee
and  $R^{He}_2(x)/R^D_2(x)$.
For the two-body channel an exact calculation was performed. In the three-body channel  average values for $k_{23}$ were inserted in Eq. (\ref{overl}).
Our preliminary results are shown in Fig. 3 and encourage us in performing the full LF calculation. 

\begin{figure}
\vspace{-0.8cm}
\centering
 \includegraphics[width=11.cm]{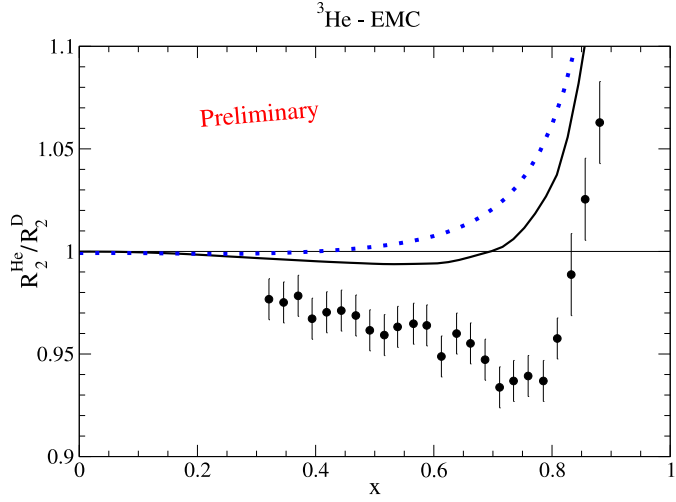}
\vspace{-0.8cm}
\caption{$^3He$ EMC effect. Solid line: result for the LF SF, with exact calculation in the 2-body channel, and  average energies in the 3-body one: $<k_{23} >$= 113.53 MeV (proton), $< k_{23} >$= 91.27 MeV (neutron), corresponding to the average kinetic energy of the intrinsic motion of the 
(23) pair in the continuum 
spectrum.
Dotted line: result 
with the approach of Ref. \cite{Sauer} for the SF. Experimental data are from Ref. \cite{Seely}.}
\label{fig1}       
\end{figure}


\section{Conclusions and Perspectives}
An investigation of SIDIS processes off $^3He$ beyond the NR, impulse approximation approach is presently being carried out.
A Generalized Eikonal Approximation has been used to deal with  the
FSI effects  and a distorted spin-dependent spectral function, still non relativistic, has been defined
\cite{Kaptari,Kaptari1}.
It has been shown that the formula (\ref{extrac}) can be safely used to obtain  both the Collins and Sivers neutron asymmetries from  the measured Collins and Sivers asymmetries of $^3He$ 
\cite{Kaptari1}.

A Poincar\'e covariant description of  A=3 nuclei, based on the LFHD, has been proposed
\cite{DPSS}. The BT construction of the Poincar\'e generators allows one to embed the successful NR phenomenology for few-nucleon systems in a Poincar\'e covariant framework. 
Then a LF SF can be defined that exactly fulfills both the normalization and the momentum sum rule.
The nucleon SF for $^3He$, has been evaluated by approximating the IF overlaps in Eq. (\ref{overl}) with their NR counterparts, calculated with the AV18 NN interaction, since it fulfills rotational and translational symmetries.

Let us stress  two important features of our LF spectral function : i) the definition of the nucleon momentum $\tilde{\bm \kappa}$ in the
intrinsic reference frame of the cluster (1,23); and ii) the use for the calculation of the  LF spectral function of the tensor product of a plane wave of momentum $\tilde{\bm \kappa}$ times the state which describes
the intrinsic motion of the fully interacting spectator subsystem. These new features
allows one  to take care of macrocausality and  to introduce a new effect of binding in the spectral
 function.

A first test of our approach is the EMC effect for $^3He$. The 2-body contribution to the nucleon SF has been calculated with the full expression, while for the 3-body contribution  average values for  $<k_{23}^2>$ have been used. In the comparison with experimental data, encouraging improvements clearly appear  with respect to the non-relativistic result.
Therefore, relativistic effects generated by the fulfillment of  Poincar\'e  covariance at the nucleus level, seem to be required to identify unambiguously new, genuine QCD phenomena inside the nucleus itself.

Our next steps will be the  full calculation of the EMC effect for $^3He$, including the exact 3-body contribution, and the introduction of the FSI  through the GEA within the LFHD.


\begin{thebibliography}{99}
 \bibitem{Qian}
	 X. Qian et al. [Jefferson Lab Hall A Collaboration],  \emph{Single Spin Asymmetries in Charged Pion Production from Semi-Inclusive Deep Inelastic Scattering on a Transversely Polarized $^3He$ Target at $Q^2$ = 1.4 - 2.7 $GeV^2$}, \emph{Phys. Rev. Lett.} {\bf 107}  (2011) 072003.
 
	\bibitem{SIDIS} H. Gao et al., \emph{Transverse spin structure of the nucleon through target single-spin asymmetry in semi-inclusive deep-inelastic (e, $e^{\prime}$ $\pi^{\pm}$ ) reaction at Jefferson Lab}, \emph{Eur. Phys. J. Plus} {\bf 126}  (2011) 2; G. Cates et al., E12-09-018, JLAB approved experiment;
 J.P. Chen et al., PR12-11-007 (Rating A), \emph{Asymmetries in Semi-Inclusive Deep-Inelastic 
 (e, e'$\pi^{\pm}$) Reactions on a Longitudinally Polarized $^3He$ Target}.
 \bibitem{06010} J.-P. Chen,  X. Jiang
and J.-C. Peng, E-06-010 Proposal to JLab-PAC29,  E. Cisbani, H. Gao and  X. Jiang, E-06-011 Proposal to JLab-PAC29, \emph{Measurement of Single Target-Spin Asymmetry in Semi-Inclusive
Pion Electroproduction on a Transversely Polarized $^3He$ Target}.

\bibitem{BARONE} V. Barone, F. Bradamante and A. Martin, \emph{Transverse-spin and transverse-momentum effects in high-energy
processes}, \emph{Prog. Part. Nucl. Phys.} {\bf 65}  (2010) 267.



\bibitem{Kaptari} L. Kaptari, A. Del Dotto, E. Pace, M. Rinaldi, G. Salm\'e,   S. Scopetta,  \emph{Distorted spin-dependent spectral function of an A=3 nucleus and semi-inclusive deep inelastic scattering processes}, \emph{Phys. Rev. C}  {\bf 89} (2014) 035206.


\bibitem{DelDotto} A. Del Dotto, L. Kaptari, E. Pace, G. Salm\'e, S. Scopetta,  \emph{Towards an Improved Description of SiDIS by a Polarized $^3He$ Target},  \emph{Few Body Syst. } {\bf 55} (2014) 877-880.
\bibitem{Kaptari1} L. Kaptari, A. Del Dotto, E. Pace, G. Salm\'e, S. Scopetta, to be published.

\bibitem{DPSS} A. Del Dotto, E. Pace, G. Salm\'e, S. Scopetta,  \emph{Light-front spin-dependent spectral function and nucleon momentum distributions for a three-body system},  \emph{Phys. Rev. C} {\bf 95} (2017) 014001.

\bibitem{KP} B.D. Keister and W.N. Polyzou,  \emph{Relativistic Hamiltonian Dynamics in Nuclear and Particle Physics}, \emph{ Adv. Nucl. Phys.}  {\bf 21} (1991)  225-550.

\bibitem{Baka} B. Bakamjian and L.H. Thomas,  \emph{Relativistic Particle Dynamics. II*}, \emph{ Phys. Rev.} {\bf  92} (1953) 1300-1310.

\bibitem{mio} S. Scopetta,  \emph{Neutron single spin asymmetries from semi-inclusive deep inelastic scattering off transversely polarized $^3He$}.  \emph{Phys. Rev. D} {\bf{75} } (2007) 054005-1.

\bibitem{cps} C. Ciofi degli Atti, E. Pace, G. Salm\`e,  \emph {Spin dependent spectral function of $^3He$ and the asymmetry in the process polarized $^3{H}e$ (polarized {e}, e') X}, \emph {Phys. Rev. C}  {\bf 46} (1992) R1591;
 C. Ciofi degli Atti, E. Pace, G. Salm\`e, \emph {Investigation of the neutron form factors by inclusive quasielastic scattering of polarized electrons off polarized $^3He$: a theoretical overview}, \emph {Phys. Rev. C}  {\bf 51} (1995) 1108-1119.



\bibitem{Kope1} C. Ciofi degli Atti, B.Z. Kopeliovich,   \emph{Final state interaction in semiinclusive DIS off nuclei},   \emph{Eur. Phys. J.} {\bf  A17} (2003) 133-144.

 \bibitem{Ciofi} C. Ciofi degli Atti and L.  Kaptari,  \emph{Semi-inclusive deep-inelastic scattering off few-nucleon systems: Tagging the EMC effect and hadronization mechanisms with detection of slow recoiling nuclei},  \emph{Phys. Rev. C} {\bf 83} (2011) 044602-1,044602-8.


\bibitem{neutr} C. Ciofi degli Atti, S. Scopetta, E. Pace, G. Salm\'e, \emph{Nuclear effects in deep inelastic scattering of polarized electrons off polarized $^3He$ and the neutron spin structure functions}, \emph{Phys. Rev.} C  48 (1993) R968-R972.

\bibitem{pisa} A. Kievsky, M. Viviani, and S. Rosati, \emph{Study of bound and scattering states in three-nucleon systems}, \emph{Nucl. Phys. A}  {\bf 577} (1994) 511-527.

\bibitem{AV18} R.B. Wiringa, V.G.J. Stocks, and R. Schiavilla, \emph{Accurate nucleon-nucleon potential with charge-independence breaking}, \emph{Phys. Rev. C} {\bf 51}  (1995) 38-51.

\bibitem{Sauer} U. Oelfke, P. Sauer and F. Coester, \emph{Convolution models of deep inelastic scattering: the three-nucleon bound state as a test case}, \emph{Nucl. Phys.} A 518 (1990) 593-616.


\bibitem{Seely} J. Seely et al., \emph{New Measurements of the European Muon Collaboration Effect in Very Light Nuclei}.  \emph{Phys. Rev. Lett.} 103 (2009) 202301-1;202301-5.








\end{thebibliography}
\end{document}